\documentclass[journal]{IEEEtran}

\usepackage[latin1]{inputenc}
\usepackage{graphicx,textcomp}
\usepackage[cmex10]{amsmath}
\usepackage{amssymb,subfigure}
\usepackage{hhline,color,siunitx}
\usepackage{blindtext}

\newlength\figureheight
\newlength\figurewidth
\pdfoutput=1 
\usepackage{ifpdf}
\begin{document}
\newcommand{\FG}{\bf \color{red}}
  \renewcommand{\arraystretch}{1.3}

%
\title{3D Modeling of the Magnetization of Superconducting Rectangular-Based Bulks and Tape Stacks}

\author{M. Kapolka, V. M. R. Zerme{\~n}o, S. Zou, A. Morandi, P. L. Ribani, E. Pardo, F. Grilli
\thanks{Document written on September 19, 2017.}
\thanks{M. Kapolka, E. Pardo are with the Slovak Academy of Sciences, Institute of Electrical Engineering, Bratislava, Slovakia.}
\thanks{V. M. R. Zerme{\~n}o, S. Zou and F. Grilli are with the Karlsruhe institute of Technology, Institute for Technical Physics, Karlsruhe, Germany.}
\thanks{P. L. Ribani and A. Morandi are with the University of Bologna, Italy.}
\thanks{M. Kapolka and E. Pardo acknowledge the use of computing resources provided by the project SIVVP, ITMS 26230120002 supported by the Research \& Development Operational Programme funded by the ERDF, the financial support of the Grant Agency of the Ministry of Education of the Slovak Republic and the Slovak Academy of Sciences (VEGA) under contract No. 2/0126/15.}
\thanks{Corresponding author's email: francesco.grilli@kit.edu.}
}



\maketitle

\begin{abstract}
In recent years, numerical models have become popular and powerful tools to investigate the electromagnetic behavior of superconductors. One domain where this advances are most necessary is the 3D modeling of the electromagnetic behavior of superconductors. For this purpose, a benchmark problem consisting of superconducting cube subjected to an AC magnetic field perpendicular to one of its faces has been recently defined and successfully solved.
In this work, a situation more relevant for applications is investigated: a superconducting parallelepiped bulk with the magnetic field parallel to two of its faces and making an angle with the other one without and with a further constraint on the possible directions of the current. The latter constraint can be used to model the magnetization of a stack of high-temperature superconductor tapes, which are electrically insulated in one direction.
For the present study three different numerical approaches are used: the Minimum Electro-Magnetic Entropy Production (MEMEP) method, the $H$-formulation of Maxwell's equations and the Volume Integral Method (VIM) for 3D eddy currents computation. The results in terms of current density profiles and energy dissipation are compared, and the differences in the two situations of unconstrained and constrained current flow are pointed out. In addition, various technical issues related to the 3D modeling of superconductors are discussed and information about the computational effort required by each model is provided. This works constitutes a concrete result of the collaborative effort taking place within the HTS numerical modeling community and will hopefully serve as a stepping stone for future joint investigations.
\end{abstract}


\IEEEpeerreviewmaketitle


\section{Introduction}

Numerical models have become popular tools for investigating the behavior and predict the performance of high-temperature superconductor (HTS) applications~\cite{Sirois:SST15,Grilli:TAS16a}. Following this development, dedicated workshops~\cite{nummodworskshop5} and -- more recently -- a summer school~\cite{School_Modeling_2016} have been organized. A workgroup and a website have been set up to collect related publications, share model files and propose benchmarks~\cite{nummodweb}.

At the workshops it has been recognized that 2D modeling of superconductors has reached a mature status of development, with models able to handle the very large numbers of conductors in coils and magnets, and very detailed material properties, including, for example, complex angular dependencies of the critical current density on the magnetic field.
On the other hand, 3D modeling is not very widely diffused in the applied superconductivity community: several numerical approaches have been proposed and used to solve relatively simple problems~\cite{Pecher:EUCAS2003, Grilli:SST03, Campbell:SST09, Farinon:SST14, Pardo:TAS16}, but without the accuracy, benchmarking and validation with experimental data typical of their 2D counterparts.

In order to support the development and validation of efficient 3D model, a benchmark has been set: the simulation of a cube of an HTS cube subjected to an AC magnetic field. This is a situation for which analytical solutions do not exist and cannot be simplified to a 2D problem. Recent work by Pardo and Kapolka with the Minimum Electro-Magnetic Entropy Production (MEMEP) method has shown, for a homogeneous and isotropic superconductor, the existence of current paths outside the planes perpendicular to the applied field~\cite{Pardo:SST17,Pardo:JCP17}. The benchmark has been solved by three different numerical approaches, and the results published on the website~\cite{nummodweb}.

\begin{figure}[t!]
\centering
\includegraphics[width=8 cm]{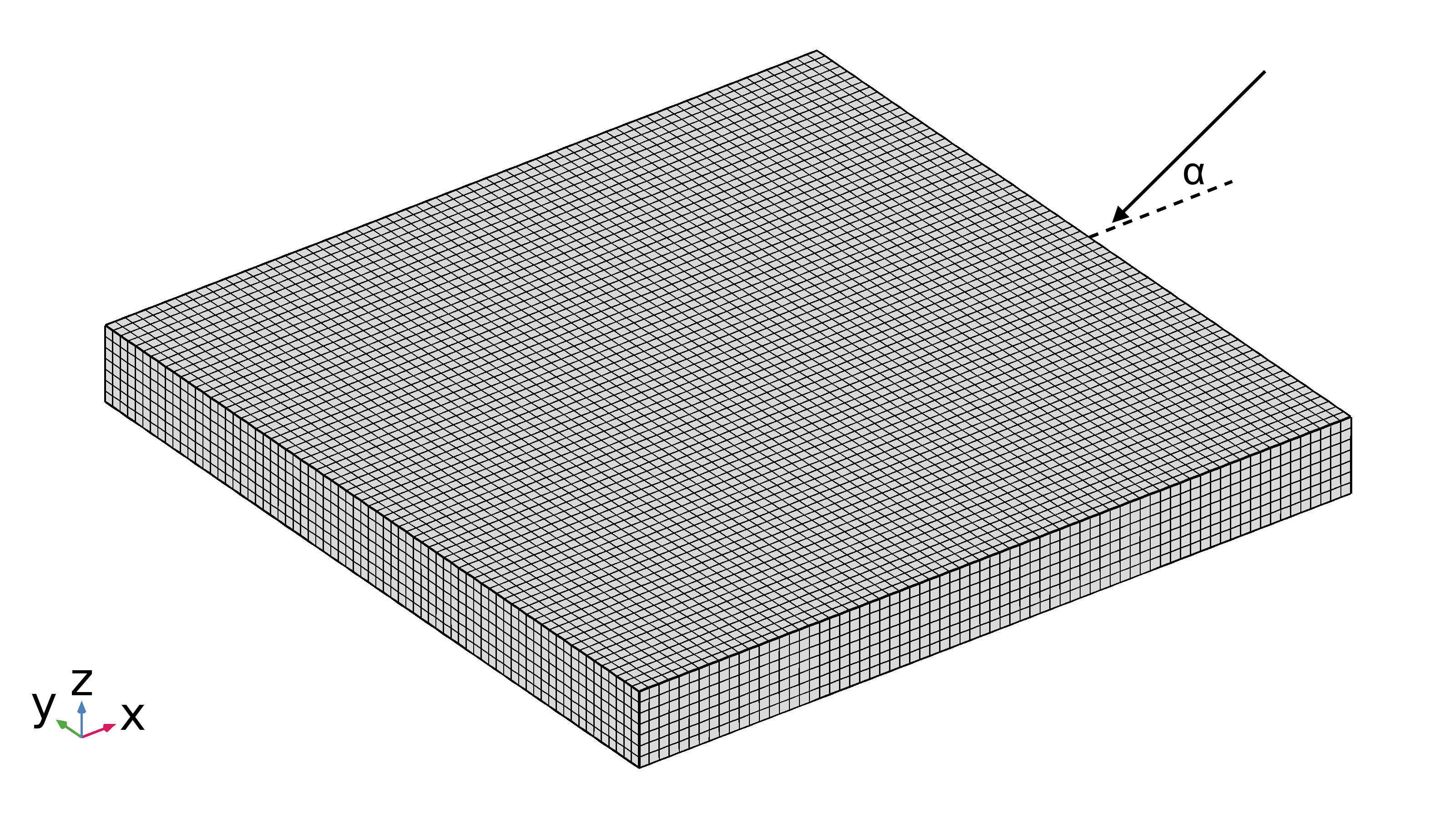}
\caption{\label{fig:gimp}Geometry of the parallelepiped under study. The external AC magnetic field is applied in the $xz$ plane, at \SI{30}{\degree} with respect to the $x$ axis. The volume is discretized in $71 \times 71 \times 7$ cells.}
\end{figure}

In this contribution, we push forward the 3D numerical modeling and investigate the magnetization of a superconducting parallelepiped subjected to an AC magnetic field making an angle with the normal to the larger surface, as shown in Fig.~\ref{fig:gimp}.
This is a situation of interest because, in addition to be a fully 3D situation for which no analytical solutions exist, it represents a configuration similar to that found in magnetization measurements of superconducting samples by VSM or SQUID magnetometers. 

In addition to a homogeneous and isotropic superconductor, we also study the case where no current can flow in the $z$ direction. This condition can be used to model different realistic cases: for example, a bulk ceramic HTS material with the $c$ axis parallel to $z$, in which, due to the strong anisotropy, no substantial current circulates in the $z$ direction. Another case of practical interest is that of stacks of HTS coated conductors used as permanent magnets in superconducting motor applications~\cite{Patel:SST13}, where the current cannot flow along $z$ due to the presence of high-resistivity layers in the stack.

For this investigation, we use different numerical approaches: in addition to the already mentioned MEMEP method, the $H$-formulation of Maxwell's equations and the Volume Integral Method (VIEM) for 3D eddy currents computation.

\section{Problem Definition}
The problem under investigation is that of a superconducting parallelepiped, with a square base of side $w=\SI{10}{\milli\meter}$ and height \SI{1}{\milli\meter}, as represented in Fig.~\ref{fig:gimp}.
The parallelepiped is subjected to a uniform external sinusoidal magnetic field in the ${xz}$ plane at an angle $\alpha=\SI{30}{\degree}$ with respect to the $x$ axis, of amplitude $B_{\rm a}=\SI{200}{\milli\tesla}$ and frequency $f=\SI{50}{\hertz}$. The superconductor is modeled as a material with non-linear resistivity
\begin{equation}\label{eq:PL}
\rho(J)=\frac{E}{E_{\rm c}}\left | \frac{J}{J_{\rm c}} \right | ^{n-1},
\end{equation}
where $J$ and $E$ are the moduli of the current density and of the electric field, respectively, $E_{\rm c}=\SI{1e-4}{\volt\per\meter}$, $J_{\rm c}=\SI{1e8}{\ampere\per\meter\squared},$ $n=25$. Under the assumption of isotropic medium, the directions of the electric field and of the current density are the same.

In addition to an isotropic resistivity (corresponding to a homogeneous and isotropic superconducting bulk), the situation where the current cannot flow in the $z$ direction is also considered. This constraint on the direction of the current flow can be used to model the magnetization of stacks of HTS coated conductors as a homogeneous bulk~\cite{Zou:Thesis17}. For brevity, the two situations will be refereed to as bulk and stack, respectively. 

The AC loss (per cycle) is calculated in two different ways~\cite{Grilli:TAS14a}:
\begin{itemize}
\item by integrating the instantaneous power dissipation $\mathbf{J}\cdot {\mathbf E}$ in the superconductor
\begin{equation}\label{eq:Q_JE}
Q_{\rm JE}=2\int_{1/2f}^{1/f} \int_{\Omega} \mathbf{J}\cdot {\mathbf E}~{\rm d}\Omega{\rm d}t,
\end{equation}
where $\Omega$ is the superconducting volume and the time interval is taken after half AC cycle;
\item by integrating the magnetization loop
\begin{equation}\label{eq:Q_MH}
Q_{\rm MH}=-\mu_0\oint m_{\rm a} {\rm d} H_{\rm a},
\end{equation}
where $m_{\rm a}$ is the magnetic moment in the direction of the applied field $H_{\rm a}$. 
\end{itemize}

\section{Description of the Numerical Models}

The Minimum Electromagnetic Entropy Production (MEMEP) model is based on a variational method. The method minimizes a certain functional~\cite{Pardo:JCP17}, which uses the effective magnetization $\mathbf T$. The $\mathbf  T$ vector is zero outside the sample, and hence the method discretizes the geometry of only the sample, which reduces the degrees of freedom and computing time. The mesh is split to three sets of sectors, which are overlapped by 1/3 of sector size. The use of sectors further reduces the computing time and prepares the program for highly efficient parallel calculation. The program is written in C++ and uses protocols like OpenMP and BoostMPI for parallel calculation in computer clusters.

The FEM model is based on the $H$-formulation of Maxwell's equations implemented in the finite-element software package Comsol Multiphysics~\cite{Brambilla:SST07,Grilli:Cryo13,Zermeno:SST13}. The superconducting parallelepiped is surrounded by a cubic air domain (side \SI{100}{\milli\meter}), on whose faces the external magnetic field is applied as a Dirichlet boundary condition. The air is modeled as a material with very large electrical resistivity (\SI{1}{\ohm\meter}). In the case of the homogenized stack, the superconductor is assigned an anisotropic resistivity, the power-law resistivity of~\eqref{eq:PL} in the $x$ and $y$ direction, and $\rho=\SI{1}{\ohm\meter}$ in the $z$ direction, so that the current cannot flow along $z$.
The superconducting and air domains are meshed with hexahedral and tetrahedral  elements, respectively. A ``jacket'' of prism elements has to be created around the superconducting domain, in order to join the other two types of mesh elements~\cite{Zou:Thesis17}.

The volume integral equation method (VIEM) is based on the ${\mathbf A}-\phi$ formulation of the eddy current problem and is implemented by means of a homemade computer program written in FORTRAN90. The superconductor domain is subdivided in a finite number of hexahedral elements. No discretization of the surrounding air is needed. Edge elements shape functions are used for relating the current density of each element to the currents through its faces~\cite{Albanese:AIEP97}. A weak solution is obtained by taking the loop integrals of the electric field, which allows eliminating the scalar electric potentials from the set of the unknowns. In essence, a circuit is associated to the superconductor domain which contains as many nodes as the number of elements and as many branches as the faces of the mesh noy lying on the boundary. The unknowns of the problem are the loop currents of the circuit. As the current of each element contributes to the magnetic vector potential at any point, a dense interaction matrix is obtained, as inherent for integral methods. Details of the model can be found in~\cite{Cristofolini:PhysC02, Morandi:Thesis04}.

\section{Results}
This section focuses on the comparison of current density distributions and AC losses in the case of bulk and the stack.

\begin{figure}[h!]
\centering
\includegraphics[width=8 cm]{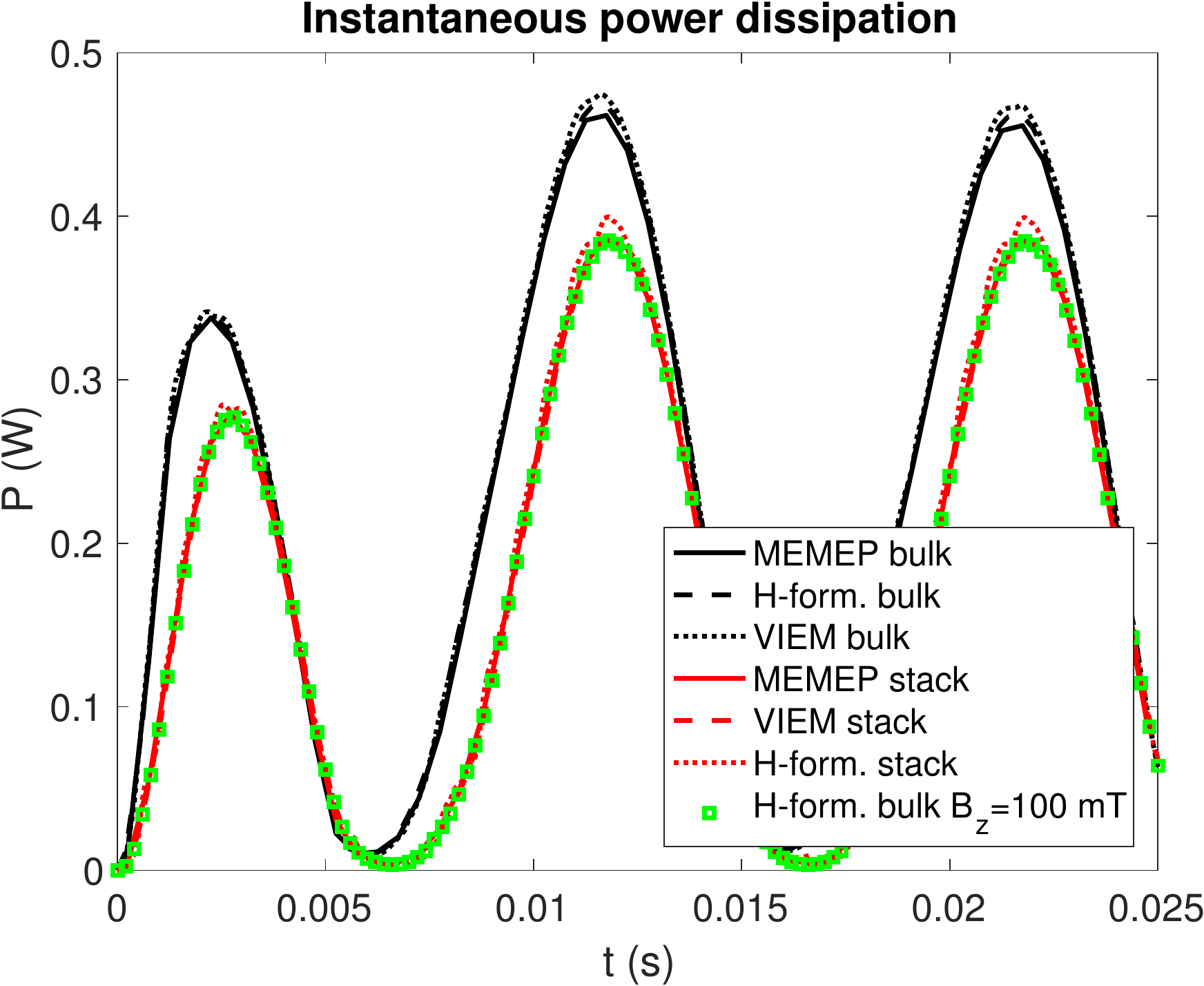}
\caption{\label{fig:Pt}Instantaneous power dissipation in the bulk (black) and stack (red), calculated with the different models: MEMEP (continuous line), $H$-formulation (dashed line) and VIEM (dotted line). The results for a bulk subjected to a field of \SI{100}{\milli\tesla} (calculated with the $H$-formulation) is also shown.}
\end{figure}

The instantaneous power loss during the first~\SI{25}{\milli\second} (rise to the peak and first cycle of the applied field) is shown in Fig.~\ref{fig:Pt}. The total energy loss during one cycle (calculated in the time interval from peak to peak of the applied field) are reported in Table~\ref{tab:Q}. The results obtained with the different models are in excellent agreement with each other. This happens for both methods used to calculate the losses, given by equations~\eqref{eq:Q_JE} and \eqref{eq:Q_MH}. The agreement between the models is consistent with the previously reported results found for the benchmark case of a cube subjected to an AC field perpendicular to one of its faces, which are fully documented in~\cite{nummodweb} (Benchmark \#5).

\begin{table}[h!]
\centering
\caption{\label{tab:Q}Comparison of AC losses (in \SI{}{\milli\joule}) in the bulk and in the stack, calculated with two different methods.}
\begin{tabular}{c c c c c}
 		& $Q_{\rm JE}$ bulk 	& $Q_{\rm MH}$  bulk	& $Q_{\rm JE}$ stack	& $Q_{\rm MH}$ stack\\    \hline
 MEMEP 	& 4.58 		& 4.62 		& 3.48 		& 3.50\\
 $H$-formulation & 4.59 	& 4.62		& 3.47 		& 3.45\\
 VIEM 		& 4.67 		& 4.70		& 3.56		& 3.56
\end{tabular}
\end{table}

\begin{figure}[h!]
\centering
\includegraphics[width=8 cm]{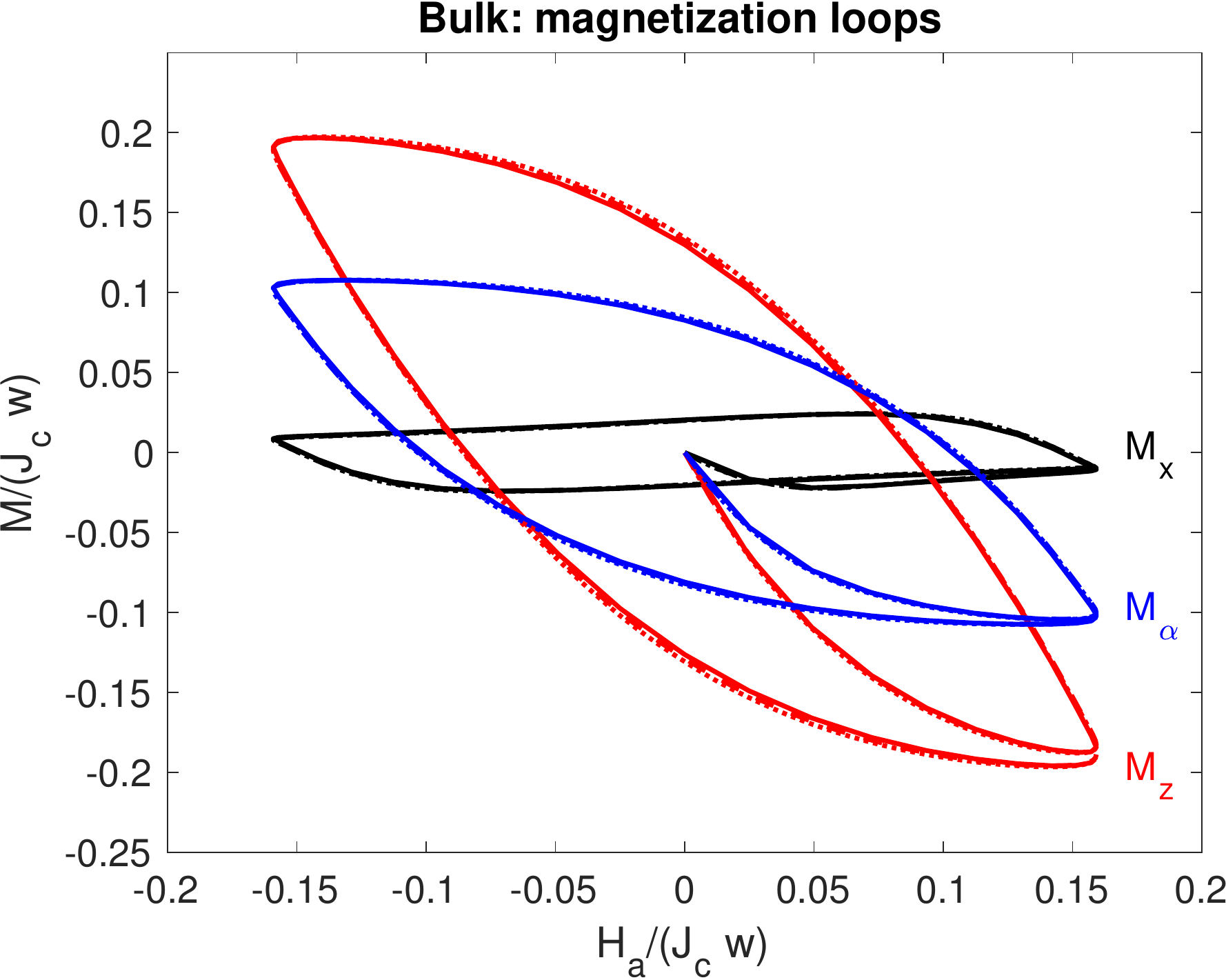}
\caption{\label{fig:MH}Magnetization loops of the bulk: the magnetization is calculated along $x$ (black), $z$ (red) and the direction of the field (blue). Each magnetization curve is calculated with the three models: MEMEP (continuous line), $H$-formulation (dashed line) and VIEM (dotted line). The results of the different models perfectly overlap.}
\end{figure}

\begin{figure}[h!]
\centering
\includegraphics[width=8 cm]{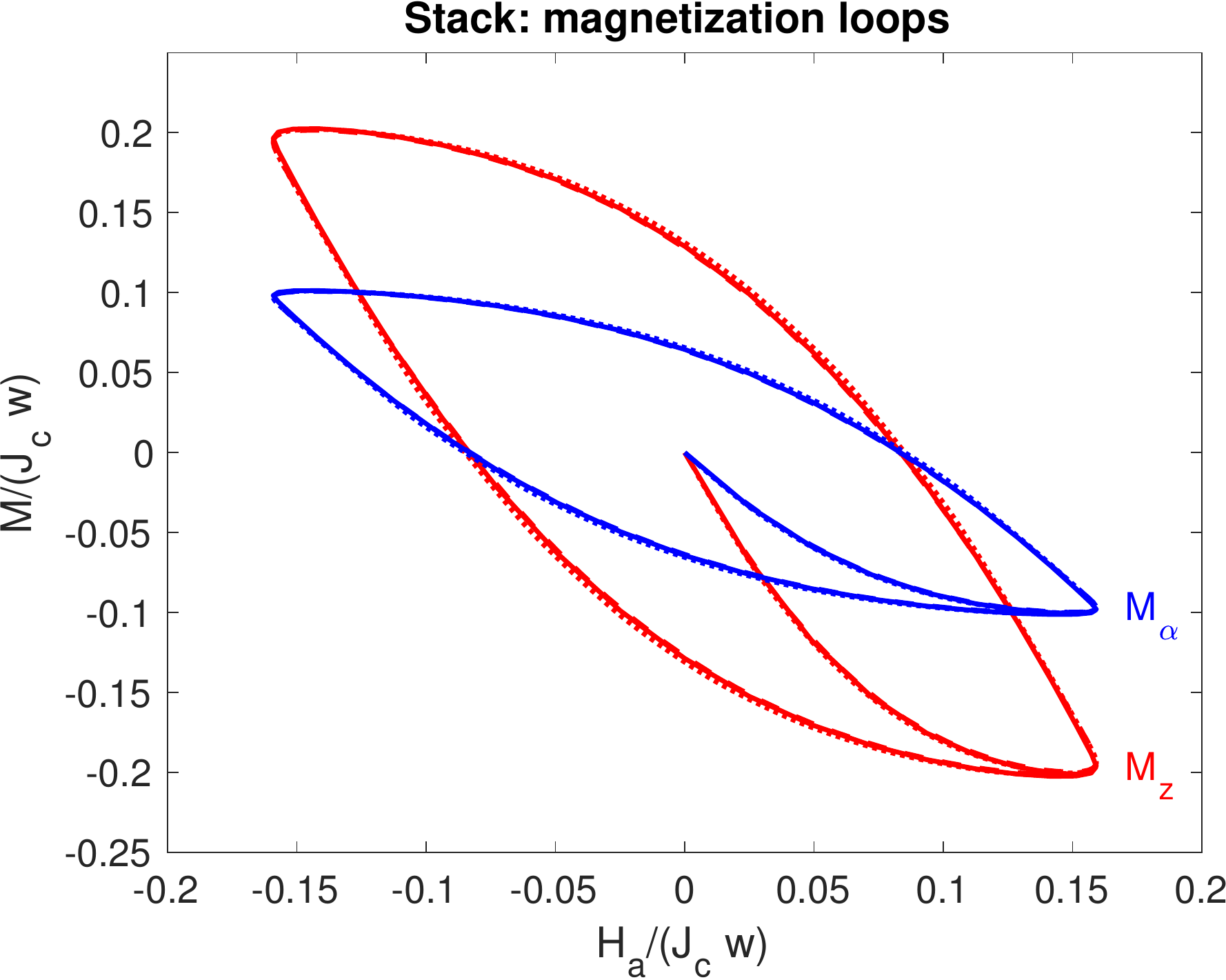}
\caption{\label{fig:MH_rho1}Magnetization loops of the stack. The magnetization is calculated along $z$ (red) and the direction of the field (blue). Each magnetization curve is calculated with the three models: MEMEP (continuous line), $H$-formulation (dashed line) and VIEM (dotted line). The results of the different models perfectly overlap.}
\end{figure}

Figures~\ref{fig:MH} and~\ref{fig:MH_rho1} show the magnetization loops in the bulk and the stack respectively. The $x$ and $z$ components of the magnetization are shown along with the component along the magnetic field direction $M_{\alpha}=M_x \cos\alpha+M_y\sin\alpha$. The data shown are obtained by the ration between the calculated total magnetic moment and the volume of the superconductor domain, normalized with respect to $J_{\rm c}w$ (with $w$ being the side of the base of the parallelepiped). The applied magnetic field appearing on the abscissae is normalized by $J_{\rm c}w$ as well, so that both quantities plotted in the graph are non-dimensional. 

Figures~\ref{fig:Pt}, \ref{fig:MH} and \ref{fig:MH_rho1} indicate that the stack has lower losses than the bulk. This is because a lower total shielding current is induced in the stack. In particular, in the stack case, the $x$ component of the magnetic field cannot induce current loops in the $yz$ plane, so the situation is similar to having just a field of amplitude $\SI{200}{\milli\tesla} \times \sin \alpha=\SI{100}{\milli\tesla}$ directed along $z$. This is confirmed by the curves of instantaneous power dissipation in Fig.~\ref{fig:Pt}.
 
In the stack, the fact of having the constraint that no current can flow in the $z$ direction leads to very different current density distributions than in the case of the bulk. All the current density distributions that follow are taken at the instant of the peak of the applied field, $t=\SI{5}{\milli\second}$.

\begin{figure}[t!]
\centering
\includegraphics[width=8 cm]{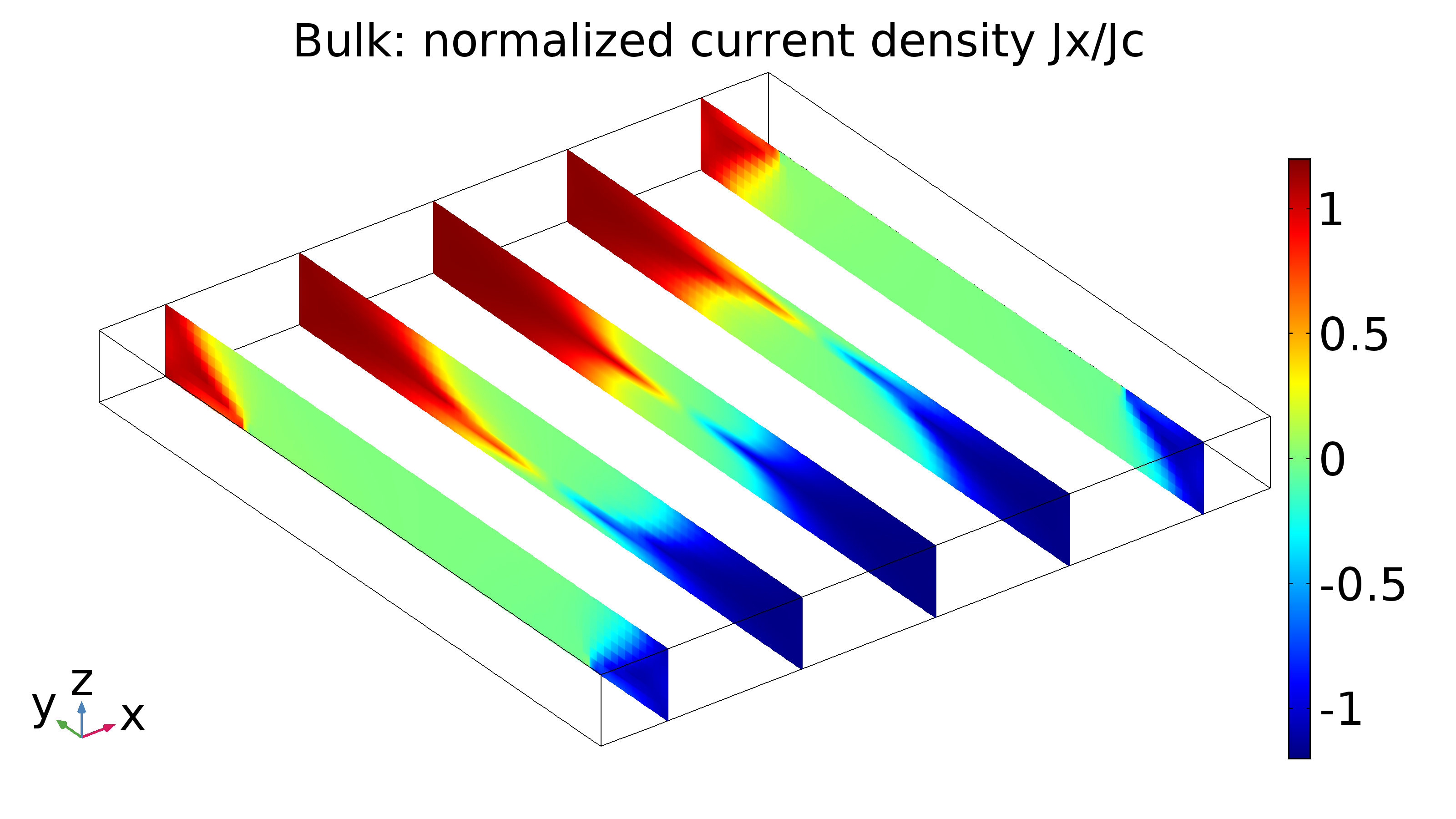}
\includegraphics[width=8 cm]{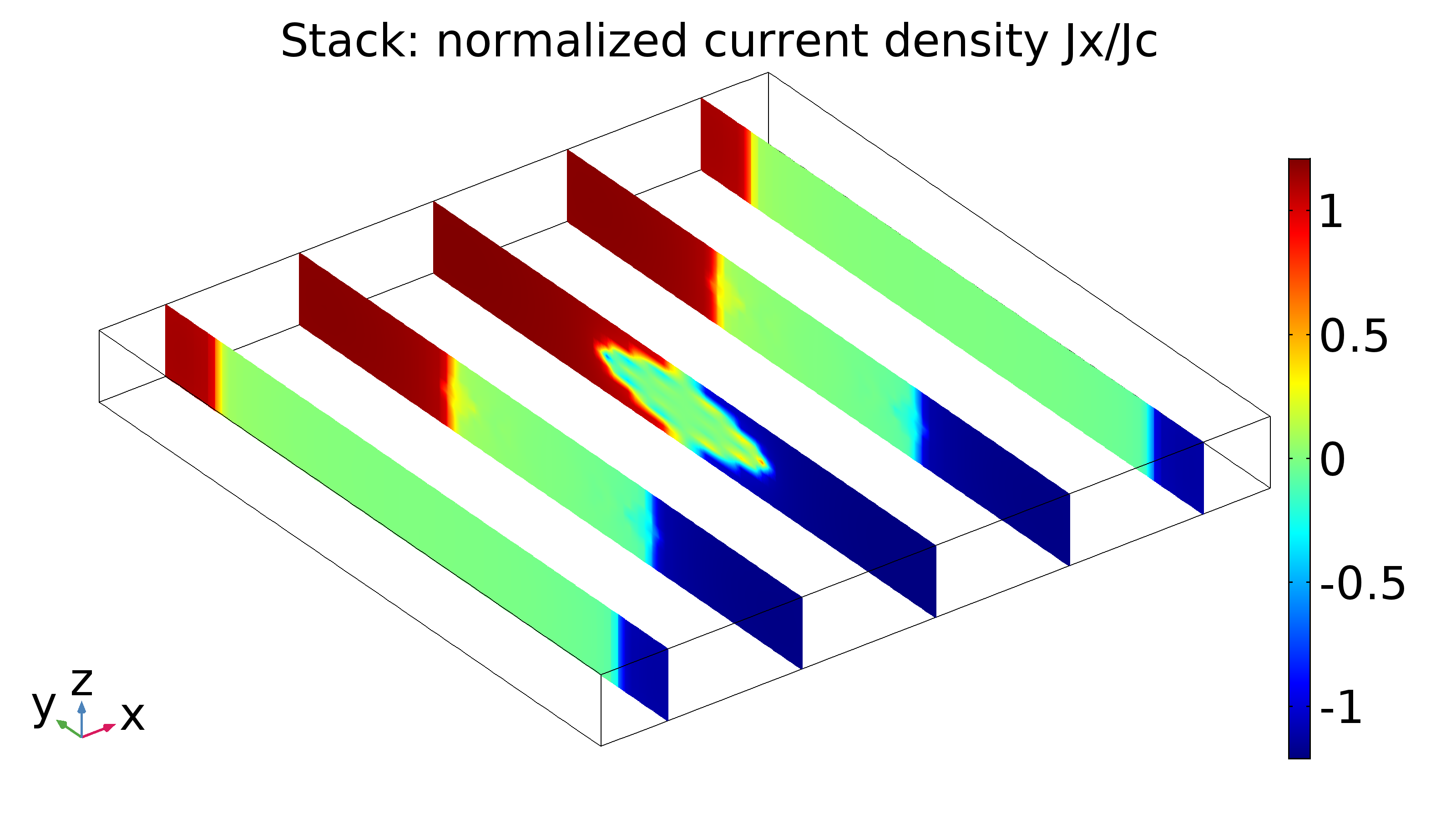}
\caption{\label{fig:Jx}Distribution of $J_x$  on five $yz$ cross-sections of  the bulk (top) and stack (bottom).}
\end{figure}

Fig.~\ref{fig:Jx} shows that, in the bulk, the pattern of the $J_x$ component is influenced by the angle of the incident magnetic field: with the exception of the mid-plane (the third of the five displayed), the $J_x$ distribution varies along $z$. In the stack, the $J_x$ distribution is perfectly symmetric. In the mid-plane, it resembles the current distribution of a stack of long tapes subjected to a field perpendicular to the flat face of the tapes. One can compare this, for example, to the 2D results shown in Fig.~2 of~\cite{Grilli:TAS07a} or Fig.~6 of~\cite{Pardo:SST12b}.

\begin{figure}[t!]
\centering
\includegraphics[width=8 cm]{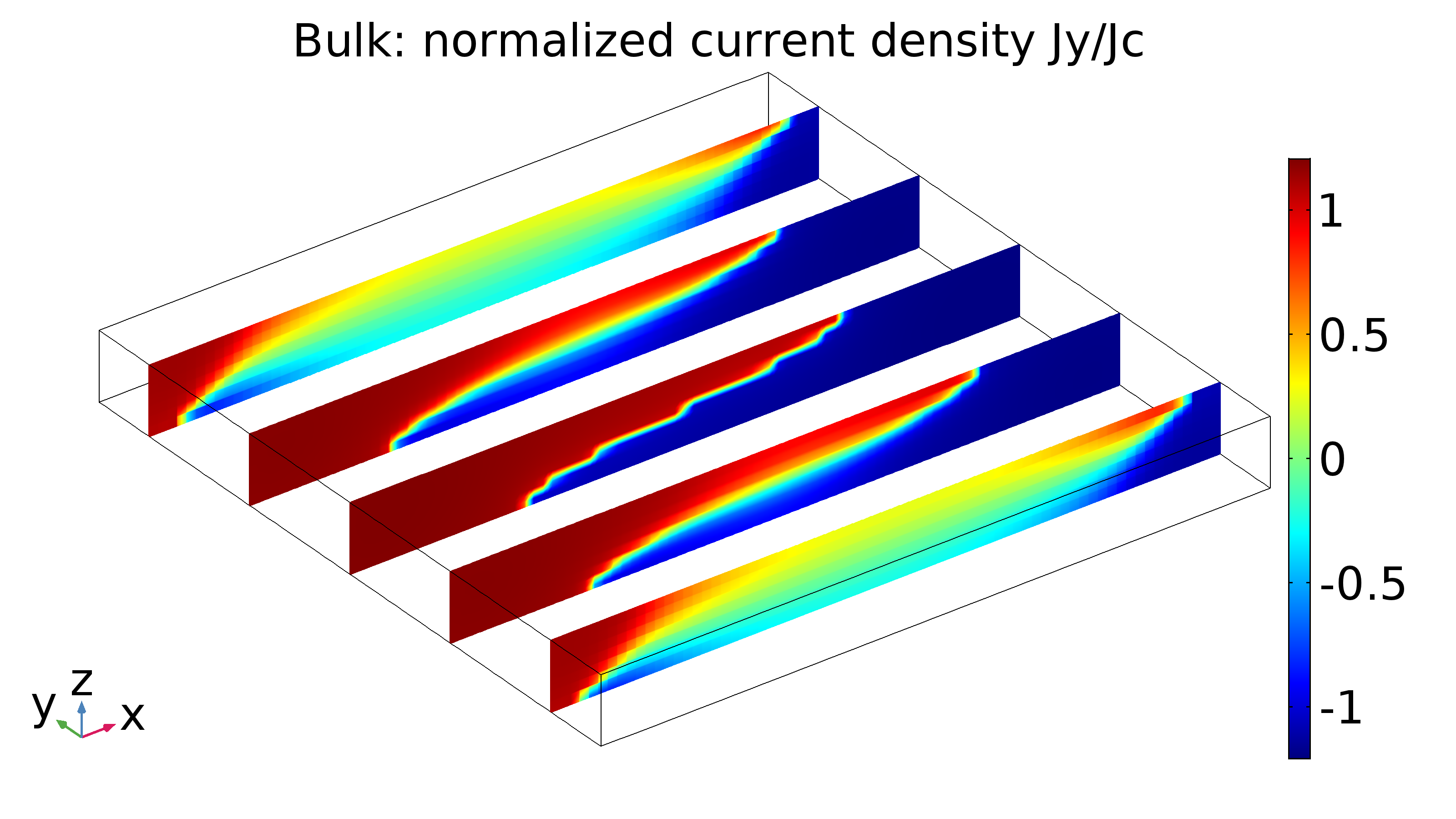}
\includegraphics[width=8 cm]{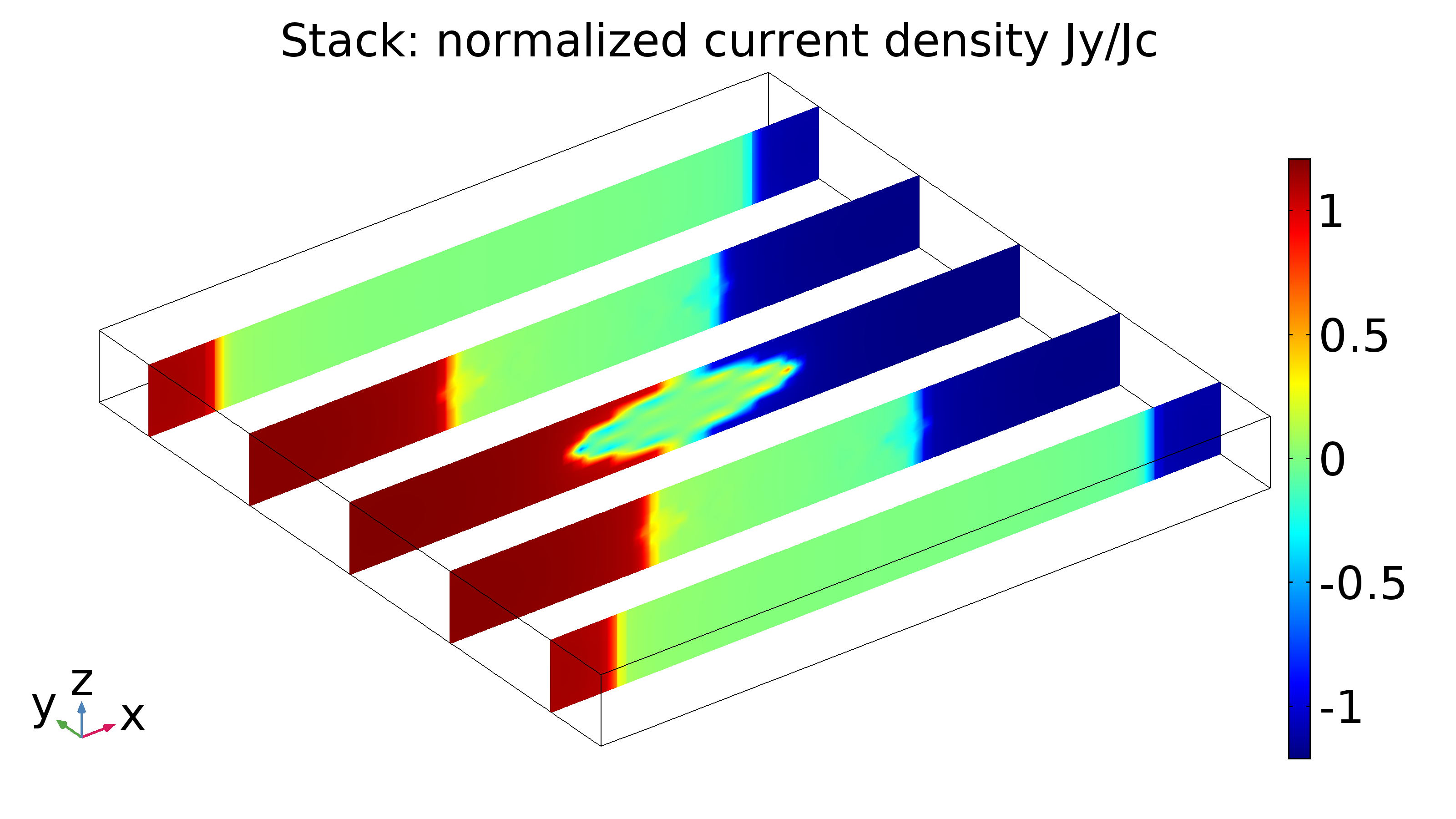}
\caption{\label{fig:Jy}Distribution of $J_y$  on five $xz$ cross-sections of  the bulk (top) and stack (bottom).}
\end{figure}

The differences between bulks and tapes are even more striking by looking at the $J_y$ component (Fig.~\ref{fig:Jy}). At the mid-plane of the bulk, the cross-section is completely saturated with positive and negative currents, because the field is large enough to fully penetrate the bulk. The line of separation between positive and negative currents is oriented along the direction of the applied field. At the mid-plane of the stack, however, the situation is completely different. The distribution is symmetric (in fact, it is identical to the $J_x$ distribution discussed above): in the stack the current cannot flow along $z$, and so only the field directed along $z$ can induce current in the the superconductor. Interestingly, the amplitude of the $z$ component is only half of that of the applied field ($\sin \alpha = 0.5$), and it is not sufficient to fully penetrate the superconductor (as, on the contrary, was the case with the bulk). As a consequence, a current-free core remains.

\begin{figure}[t!]
\centering
\includegraphics[width=8 cm]{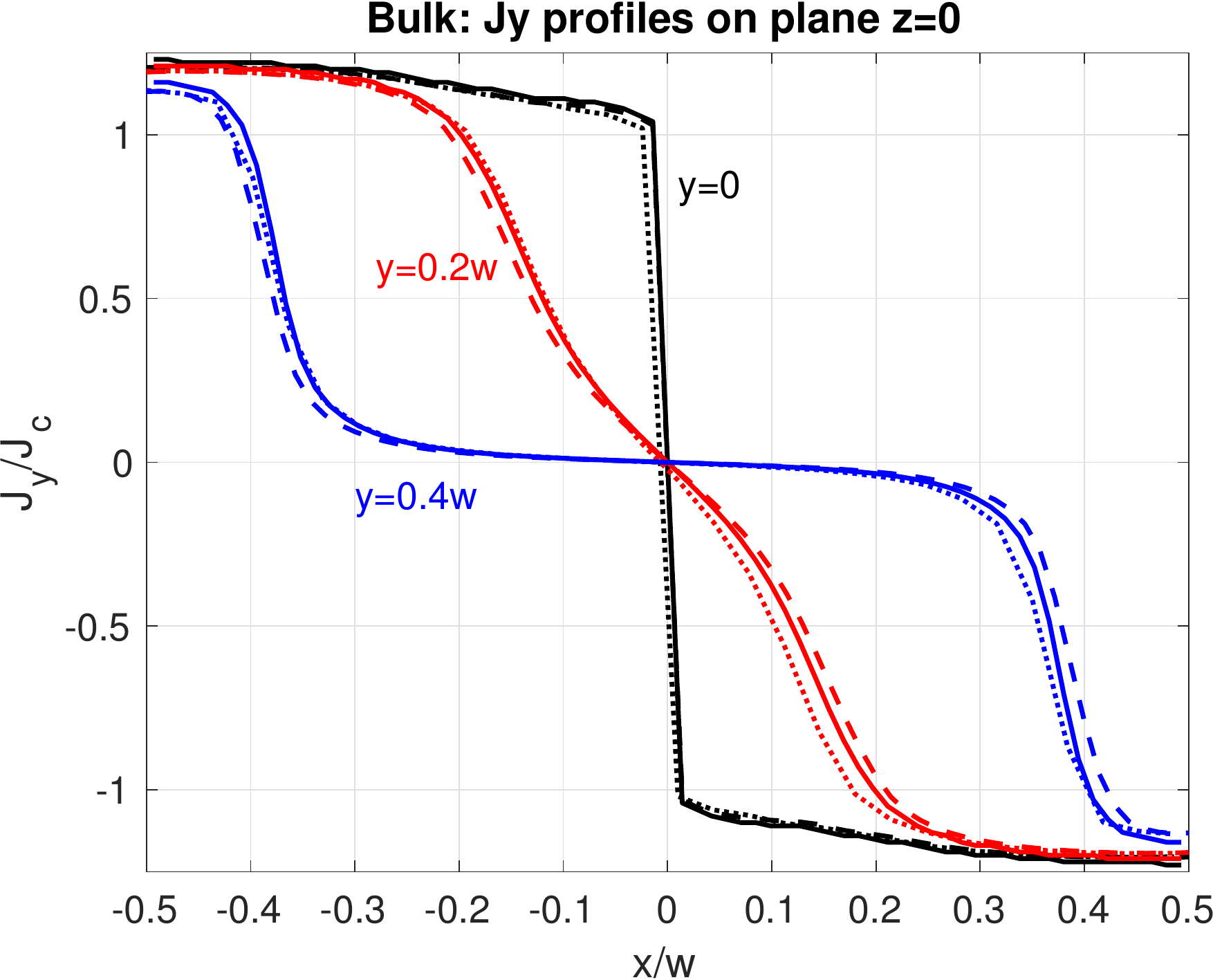}
\caption{\label{fig:Jy_profiles}Distribution of $J_y$  along three lines on the $z=0$ plane, calculated with the three models: MEMEP (continuous line), $H$-formulation (dashed line) and VIEM (dotted line).}
\end{figure}

A more detailed view of the $J_y$ distribution in the bulk is given in Fig.~\ref{fig:Jy_profiles}, which shows the profiles  along three particular lines on the plane $z=0$: the excellent agreement between the models is demonstrated not only for integral quantities like AC loss and magnetization, but also for local current density profiles.

\begin{figure}[t!]
\centering
\includegraphics[width=8 cm]{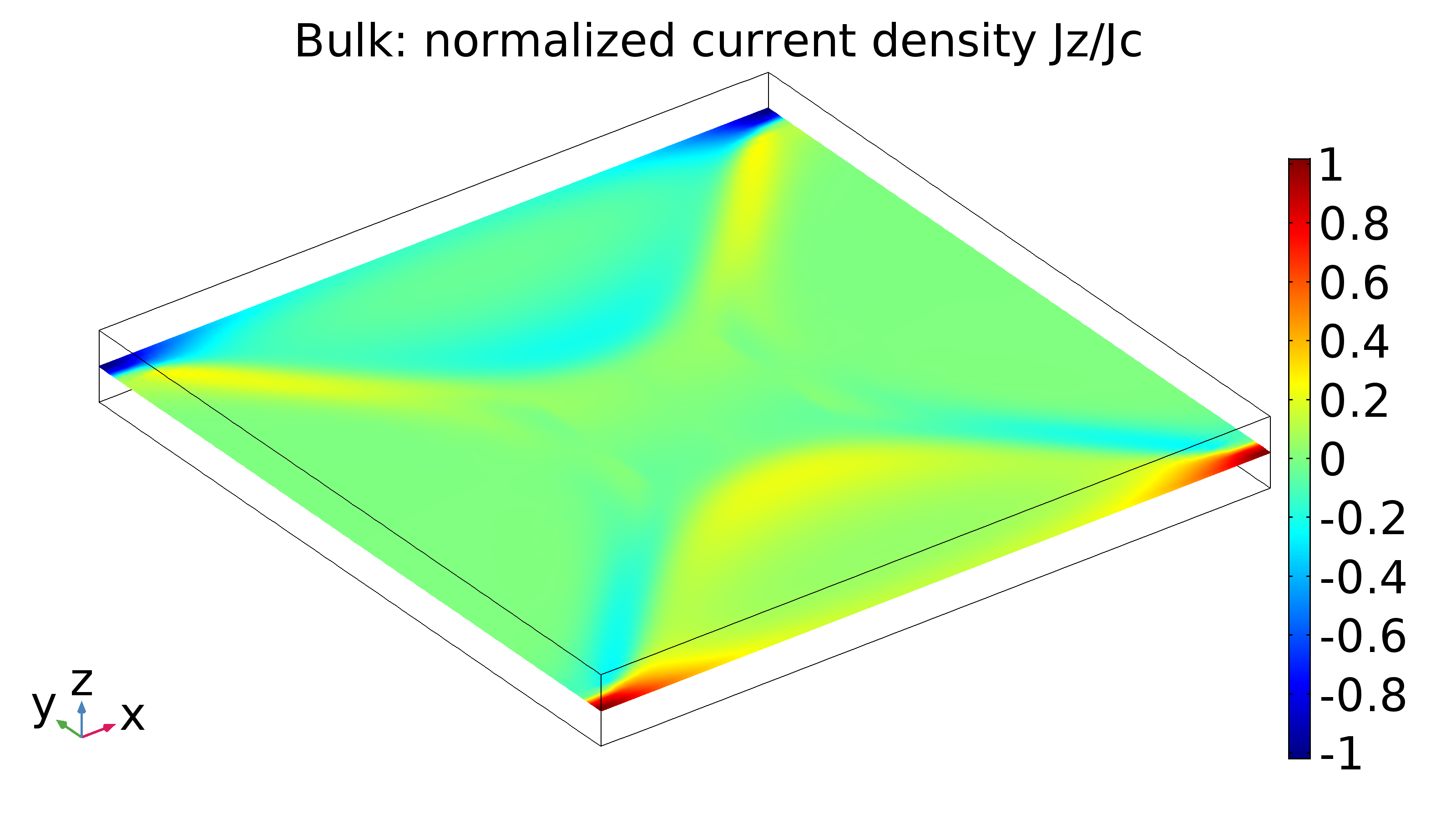}
\caption{\label{fig:Jz_bulk}Distribution of $J_z$  on the plane $z=0$ of the bulk.}
\end{figure}

Finally, Fig.~\ref{fig:Jz_bulk} shows the $J_z$ component in the bulk on the plane $z=0$. With the exception of some small  concentrations in the corners, these currents are small (about \SI{30}{\percent} of $J_{\rm c}$ at most), but not negligible.

\begin{table}[h!]
\centering
\caption{\label{tab:tcalc}Computation times for bulk and stacks for different geometry discretizations.}
\begin{tabular}{c c c}
Bulk 		& $24 \times 24 \times 8$ &  $71 \times 71 \times 7$	\\    \hline
 MEMEP$^1$ 	& 8 h 		& 7.7 d 	\\
 $H$-formulation$^2$ & 13 h 	& 4.2 d	\\
 VIM$^3$ 		& 23 h 		& --	\\ \hline
Stack 		& $24 \times 24 \times 8$ &  $71 \times 7 \times 7$	\\    \hline
 MEMEP$^1$ 	& 2.2 h		& 6.0 d 	\\
 $H$-formulation$^2$ & 15 h 	& 4.1 d	\\
 VIEM$^3$ 		& 9 h 		& --	\\
 \multicolumn{3}{l}{$^1$ Intel\textsuperscript{\textregistered} Core\textsuperscript{TM} i7-4771 CPU 16 GB RAM OS: Ubuntu 16.04LTS} \\
 \multicolumn{3}{l}{$^2$ Intel\textsuperscript{\textregistered} Core\textsuperscript{TM} i7-4960 CPU 64 GB RAM OS: Windows 7} \\
 \multicolumn{3}{l}{$^3$ Intel\textsuperscript{\textregistered} Core\textsuperscript{TM} i5-3570 CPU 8   GB RAM OS: Windows10}
 \end{tabular}
\end{table}

Table~\ref{tab:tcalc} lists typical computation times for simulating the two problems (1.25 AC cycles) with the different models and two different discretizations of the superconducting volume. The MEMEP method can take good advantage of parallelization. On a 6-node cluster, the computing times for the $71 \times 71 \times 7$ discretization of the bulk and stack reduce to 2.5 and 1 day, respectively.

\section{Conclusion}
This work presented the investigation of the magnetization of superconducting rectangular-based bulks and tape stacks by 3D numerical simulations. Three different numerical approaches were compared and an excellent agreement between them was found.
The magnetization loops, the AC loss and the current patterns caused by an external magnetic field making an angle with the normal to the larger surface of the geometries under study are different in the case of bulks and stacks, due to the fact that in the stack geometry the current cannot flow in the direction perpendicular to the faces of the tapes.
The computation times strongly depend on the utilized geometry discretization. For the considered problem, a discretization of less than 5000 cells is enough to obtain accurate results in the reasonable time of a few hours on standard desktop workstations.
This works constitutes a tangible result of the collaborative effort taking place within the HTS numerical modeling community and will hopefully serve as a stepping stone for future joint investigations.

%


\ifCLASSOPTIONcaptionsoff
  \newpage
\fi

\end{document}